\documentclass[conference]{IEEEtran}
\IEEEoverridecommandlockouts

\usepackage{cite}
\usepackage{amsmath,amssymb,amsfonts}
\usepackage{algorithmic}
\usepackage{graphicx}
\usepackage{textcomp}
\usepackage{xcolor}
\usepackage{tabularx}
\usepackage{booktabs}
\usepackage{siunitx}
\usepackage{ragged2e}
\usepackage{caption}
\usepackage{tikz}
\usepackage{multirow}

\newcommand{\rot}[1]{\rotatebox{60}{#1}}
\def\BibTeX{{\rm B\kern-.05em{\sc i\kern-.025em b}\kern-.08em
    T\kern-.1667em\lower.7ex\hbox{E}\kern-.125emX}}
\begin{document}

\newcolumntype{C}{>{\centering\arraybackslash}X}

\title{Partial multivariate transformer as a tool for cryptocurrencies time series prediction}

\author{\IEEEauthorblockN{1\textsuperscript{st} Andrzej Tokajuk}
\IEEEauthorblockA{\textit{Faculty of Electronics and Information Technology} \\
\textit{Warsaw University of Technology}\\
Warsaw, Poland \\
andrzej.tokajuk.stud@pw.edu.pl}
\and
\IEEEauthorblockN{2\textsuperscript{nd} Jarosław A. Chudziak}
\IEEEauthorblockA{\textit{Faculty of Electronics and Information Technology} \\
\textit{Warsaw University of Technology}\\
Warsaw, Poland \\
jaroslaw.chudziak@pw.edu.pl}
}

\maketitle

\begin{abstract}

Forecasting cryptocurrency prices is hindered by extreme volatility and a methodological dilemma between information-scarce univariate models and noise-prone full-multivariate models. This paper investigates a partial-multivariate approach to balance this trade-off, hypothesizing that a strategic subset of features offers superior predictive power. We apply the Partial-Multivariate Transformer (PMformer) to forecast daily returns for BTCUSDT and ETHUSDT, benchmarking it against eleven classical and deep learning models. Our empirical results yield two primary contributions. First, we demonstrate that the partial-multivariate strategy achieves significant statistical accuracy, effectively balancing informative signals with noise. Second, we experiment and discuss an observable disconnect between this statistical performance and practical trading utility; lower prediction error did not consistently translate to higher financial returns in simulations. This finding challenges the reliance on traditional error metrics and highlights the need to develop evaluation criteria more aligned with real-world financial objectives.

\end{abstract}


\section{Introduction}

The cryptocurrency market is a significant and challenging domain for financial analysis. These digital assets operate within an ecosystem defined by extreme volatility, pronounced non-stationarity, and a substantial speculative component \cite{doumenis2021critical, Bohme2015}. The central challenge is compounded by a crucial observation: cryptocurrency time series, when analyzed univariately, often exhibit characteristics that closely resemble a random walk \cite{puoti2024quantifying}. This statistical property suggests that historical prices alone possess limited predictive power, a notion that necessitates the inclusion of external information to gain any predictive edge.

This need for external data, however, exposes a core methodological dilemma in modern forecasting paradigms when applied to this noisy and complex ecosystem \cite{sonkavde2023forecasting}. On one hand, univariate models, which rely merely on past price data, are information-scarce and fail to capture the impact of complex market drivers \cite{miller2021univariate}. On the other hand, full-multivariate models—which incorporate a rich set of external features—perform well in lower-volatility equity markets \cite{szydlowski2024toward} but their high dimensionality still amplifies noise from irrelevant signals and leads to severe overfitting \cite{fernandez2024taming, hajek2023well}.

This paper contributes to the cryptocurrency forecasting literature by proposing and rigorously evaluating the Partial Multivariate Transformer (PMformer) \cite{xue2025pmformer}, a model designed to operate effectively between these extremes by strategically utilizing subsets of features. We seek to determine if a partial-multivariate model, by incorporating a strategic subset of external features, can achieve higher prediction accuracy and better trading results than both univariate and full-multivariate approaches in the cryptocurrency domain.

\section{Related Work}

Algorithmic trading applies computational models to execute strategies across markets \cite{Johnson2010AlgorithmicT,chan2013algorithmic}. In cryptocurrencies, early studies frequently documented dynamics close to a random walk \cite{urquhart2016inefficiency}, making univariate forecasting with ARIMA/GARCH or exponential smoothing unreliable \cite{katsiampa2017volatility}. This motivates moving beyond price-only inputs while acknowledging the domain’s volatility and regime shifts.

Deep sequence models, particularly Transformer variants, have been adapted to financial time series \cite{banka2025applying} and integrated into decision pipelines such as multi-agent systems that blend heuristics with neural predictors \cite{chudziak2024elliottagents}. These approaches capture long-range dependencies, yet consistently profitable prediction remains elusive; model capacity alone does not resolve nonstationarity, and trading-aware evaluation is essential. Moreover, naively enlarging the feature set can invite high variance and spurious correlations in noisy markets \cite{hajek2023well}.

Consequently, \emph{feature selection} is central to robust design. Classical methods span filters, wrappers, and embedded regularization—e.g., correlation or mutual-information filters \cite{dash1997feature}, recursive feature elimination \cite{kohavi1997wrappers}, and Lasso-style sparsity \cite{tibshirani1996regression}, with surveys emphasizing gains in stability and interpretability \cite{guyon2003introduction,liu2007computational}. Partial-multivariate architectures operationalize this principle by training on compact, informative subsets rather than the entire feature space \cite{lee2024partial}. The \emph{Partial Multivariate Transformer} (PMformer) instantiates this idea and has reported strong long-horizon accuracy with fewer parameters than full multivariate Transformers \cite{xue2025pmformer}.

Our work extends this line to cryptocurrency price prediction, benchmarking PMformer against eleven baselines and examining how partial feature inclusion affects both statistical error and trading performance.

\begin{figure*}[h!]
    \centering
    \includegraphics[width=\textwidth]{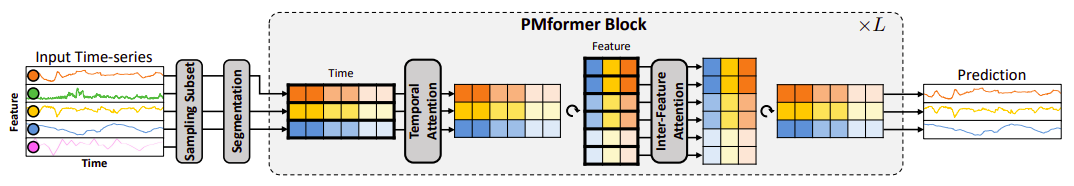}
    \caption{PMformer architecture\cite{lee2024partial}.} 
    \label{fig:pmformer-architecture}
\end{figure*}

\section{Methodology}
Our forecasting approach is centered on the Partial Multivariate Transformer (PMformer), a model that learns from strategically sampled subsets of features to predict time series \cite{lee2024partial,xue2025pmformer}. Its architecture uses a dual-attention mechanism to model both temporal and cross-sectional dependencies (Fig.~\ref{fig:pmformer-architecture}).

\subsection{Partial Multivariate Modeling}
We define the forecasting task as predicting the next value for a subset of features $\mathcal F\subset\{1,\dots,D\}$ of size $S$, where $1<S<D$. Given the historical data $X_{0:T,\mathcal F}\in\mathbb{R}^{T\times S}$, a single network $f_\theta$ outputs the prediction $\hat{x}_{T+1,\mathcal F}$:
\begin{equation}
    \hat{x}_{T+1,\mathcal F} \;=\; f_\theta\!\left(X_{0:T,\mathcal F},\,\mathcal F\right).
    \label{eq:objective}
\end{equation}
The model is trained by minimizing the Mean Squared Error (MSE) loss, averaged over time and the randomly partitioned feature subsets used in each minibatch:
\begin{equation}
    \mathcal{L}(\theta) \;=\; \mathbb{E}_{t,\mathcal F}\big[\,\|\hat{x}_{t+1,\mathcal F}-x_{t+1,\mathcal F}\|^2\,\big].
    \label{eq:loss}
\end{equation}
This training strategy on varied subsets acts as a form of regularization, improving generalization by preventing overfitting to any fixed set of feature interactions \cite{lee2024partial,xue2025pmformer}.

\subsection{PMformer Architecture}
The PMformer model processes the input subsets through three stages:

\textbf{Tokenization and Embeddings.} An input slice $X_{t-w+1:t,\mathcal F}$ is projected into tokens, which are then augmented with learnable time-step ($E^{\text{time}}$) and feature-identity ($E^{\text{feat}}$) embeddings.
\begin{equation}
    Z_0 \;=\; E_x\,X_{t-w+1:t,\mathcal F} \;+\; E^{\text{time}} \;+\; E^{\text{feat}}.
    \label{eq:tokens}
\end{equation}

\textbf{Dual-Attention Encoder.} A stack of $L$ encoder blocks processes the tokens. Each block sequentially applies temporal self-attention (\textsc{TA}) within each feature and feature-wise self-attention (\textsc{FA}) across the features.
\begin{equation}
    H^{(l)} \;=\; H^{(l-1)} \;+\; \mathrm{MLP}\!\Big(\mathrm{FA}\big(\mathrm{TA}(H^{(l-1)})\big)\Big).
    \label{eq:block}
\end{equation}

\textbf{Decoder.} A simple linear head maps the final representation $H^{(L)}$ to the prediction $\hat{x}_{T+1,\mathcal F}$.

\section{Datasets and Feature Engineering}

We use daily Bitcoin (BTCUSDT) and Ethereum (ETHUSDT) data from Binance from \mbox{Oct.\ 5, 2017 to May 20, 2025}. The raw dataset includes OHLC prices, base and quote volumes, and trade count. Extreme price movements are retained, as they represent genuine market events in crypto’s high-volatility regime.

The prediction target is the next-day logarithmic return,
$r_t=\ln(P_t/P_{t-1})$, a standard variance-stabilizing transform for financial series \cite{tsay2005analysis}. Inputs include raw OHLC, volume, and trade count augmented with widely used technical indicators \cite{kaufman1995smarter,murphy1999technical}: SMA(50), EMA(21), RSI(14), CCI(20), ATR(14), and MACD(12,26,9) (line, signal, histogram). This compact set captures trend, momentum, and volatility while remaining compatible with partial-multivariate training.

All numeric features are scaled to $[0,1]$ by Min–Max normalization \cite{goodfellow2016deep}; scaling parameters are fit on the training split and applied to the validation/test set to prevent leakage. The dataset is split chronologically into 70\% train, 20\% validation, and 10\% test. Because PMformer trains on sampled feature subsets, this acts as an implicit, adaptive feature selection step that curbs variance without exhaustive multivariate inputs \cite{guyon2003introduction}.

\begin{table*}[t]
\centering
\caption{Optimal Hyperparameter Configurations for BTCUSDT Dataset}
\label{tab:hyperparameters_btc}
\small
\setlength{\tabcolsep}{4pt}
\begin{tabular}{l|l|c|c|c|c|c|c|c|c|c|l}
\toprule
\textbf{Category} & \textbf{Model} & \textbf{LR ($\times10^{-4}$)} & \textbf{BS} & \textbf{SL} & \textbf{LL} & \textbf{Layers} & \textbf{Dim} & \textbf{H} & \textbf{d\_ff} & \textbf{D} & \textbf{Specifics} \\
\midrule \midrule
\textit{Partial-m.} & PMformer & 1.06 & 128 & 48 & 12 & $e{=}4, d{=}3$ & 64 & 16 & 128 & 0.7 & \\
\midrule
\multirow{2}{*}{\textit{Univariate}} & ARIMA & - & - & - & - & - & - & - & - & - & $p{=}2, d{=}0, q{=}0$ \\
& LSTM & 3.82 & 32 & 128 & 96 & 1 & 64 & - & - & 0.2 & \\
\midrule
\multirow{8}{*}{\textit{Multivariate}} & LSTM & 4.58 & 32 & 128 & 96 & 2 & 256 & - & - & 0.1 & \\
& Transformer & 7.61 & 128 & 192 & 48 & $e{=}2, d{=}3$ & 128 & 2 & 512 & 0.3 & \\
& Autoformer & 6.70 & 64 & 96 & 24 & $e{=}2, d{=}1$ & 128 & 4 & 32 & 0.3 & $m_{a}{=}36$ \\
& Informer & 5.95 & 32 & 192 & 96 & $e{=}4, d{=}1$ & 512 & 4 & 512 & 0.4 & \\
& FEDformer & 0.63 & 128 & 96 & 24 & $e{=}3, d{=}3$ & 64 & 16 & 128 & 0.3 & $m_{a}{=}48, v{=}Wavelets, m{=}36, m_s{=}l$ \\
& PatchTST & 0.018 & 64 & 192 & 12 & $e{=}3$ & 64 & 8 & 128 & 0.7 & $fc_{dr}{=}0.3, h_{dr}{=}0.1, p_{l}{=}16, str{=}24$ \\
& iTransformer & 3.42 & 128 & 48 & 12 & $e{=}4, d{=}3$ & 32 & 2 & 64 & 0.7 & \\
& DLinear & 7.33 & 64 & - & - & - & - & - & - & - & $individual{=}1$ \\
\bottomrule
\end{tabular}
\end{table*}

\begin{table*}[t]
\centering
\caption{Optimal Hyperparameter Configurations for ETHUSDT Dataset}
\label{tab:hyperparameters_eth}
\small
\setlength{\tabcolsep}{4pt}
\begin{tabular}{l|l|c|c|c|c|c|c|c|c|c|l}
\toprule
\textbf{Category} & \textbf{Model} & \textbf{LR ($\times10^{-4}$)} & \textbf{BS} & \textbf{SL} & \textbf{LL} & \textbf{Layers} & \textbf{Dim} & \textbf{H} & \textbf{d\_ff} & \textbf{D} & \textbf{Specifics} \\
\midrule \midrule
\textit{Partial-m.} & PMformer & 3.76 & 64 & 192 & 96 & $e{=}3, d{=}3$ & 512 & 16 & 512 & 0.4 & \\
\midrule
\multirow{2}{*}{\textit{Univariate}} & ARIMA & - & - & - & - & - & - & - & - & - & $p{=}2, d{=}0, q{=}0$ \\
& LSTM & 1.10 & 32 & 128 & 96 & 3 & 256 & - & - & 0.2 & \\
\midrule
\multirow{8}{*}{\textit{Multivariate}} & LSTM & 3.35 & 64 & 128 & 96 & 3 & 64 & - & - & 0.7 & \\
& Transformer & 2.67 & 64 & 192 & 48 & $e{=}4, d{=}2$ & 32 & 16 & 512 & 0.2 & \\
& Autoformer & 2.51 & 32 & 120 & 96 & $e{=}4, d{=}2$ & 32 & 8 & 128 & 0.7 & $m_{a}{=}24$ \\
& Informer & 2.84 & 64 & 192 & 96 & $e{=}4, d{=}1$ & 256 & 8 & 32 & 0.4 & \\
& FEDformer & 6.26 & 64 & 48 & 48 & $e{=}4, d{=}3$ & 64 & 8 & 128 & 0.7 & $m_{a}{=}37, v{=}Fourier, m{=}36, m_s{=}r$ \\
& PatchTST & 0.91 & 32 & 192 & 96 & $e{=}3$ & 32 & 4 & 512 & 0.7 & $fc_{dr}{=}0.2, h_{dr}{=}0.1, p_{l}{=}8, str{=}8$ \\
& iTransformer & 2.48 & 64 & 96 & 96 & $e{=}4, d{=}1$ & 64 & 4 & 512 & 0.4 & \\
& DLinear & 8.76 & 64 & - & - & - & - & - & - & - & $individual{=}1$ \\
\bottomrule
\end{tabular}
\end{table*}

\section{Experiments}

To empirically validate our research hypotheses, we performed a comprehensive experimental study. PMformer is benchmarked against 11 baselines—covering statistical, recurrent, and Transformer-style models, which underwent Bayesian hyper-parameter tuning. Their performance was assessed with both conventional forecasting metrics and domain-specific trading measures.

\subsection{Baseline Models}
To rigorously evaluate PMformer's performance, we benchmarked it against a comprehensive suite of models. We included a \textbf{Previous Result (Naive)} forecast as a simple, no-change benchmark. For classical statistical methods, we used an \textbf{ARIMA} model \cite{box2015time}. Foundational deep learning approaches were represented by both a \textbf{multivariate LSTM} and a \textbf{univariate LSTM} \cite{hochreiter1997long}, allowing for a direct comparison of information usage.

The core of our benchmark suite consisted of various Transformer-based architectures, including the original \textbf{Transformer} adapted for time series \cite{vaswani2017attention}, efficiency-focused variants like \textbf{Informer} \cite{zhou2021informer} and \textbf{FEDformer} \cite{zhou2022fedformer}, the decomposition-based \textbf{Autoformer} \cite{wu2021autoformer}, the patch-based \textbf{PatchTST} \cite{nie2022time}, the inverted \textbf{iTransformer} \cite{liu2023itransformer}, and the simple yet powerful linear model, \textbf{DLinear} \cite{zeng2023transformers}. This selection ensures that PMformer is compared against a representative set of alternatives.

\subsection{Training and Hyperparameters}
All models were trained on an NVIDIA T4 GPU. We run 100 Bayesian trials per model, selecting by validation MSE; each trial is averaged over three seeds (42, 1337, 2025).

Across all models we explored a single grid: learning rate \textbf{LR} was drawn log-uniformly from $10^{-5}$ to $10^{-3}$; batch size \textbf{BS}$\in\{32,64,128\}$; dropout \textbf{DR}$\in\{0.1,0.2,0.3,0.4,0.7\}$; sequence length \textbf{SL}$\in\{24,48,96,192\}$; and label length \textbf{LL}$\in\{12,24,48,96\}$ with the constraint LL\,$<$\,SL.  Architectural degrees of freedom comprised the hidden dimension \textbf{Dim}$\equiv d_{\text{model}}$ and feed-forward width $d_{\text{ff}}$, both chosen from \{32,64,128,256,512\}; encoder layers \textbf{L}$_e\in\{1,2,3,4\}$ and decoder layers \textbf{L}$_d\in\{1,2,3\}$ (or total recurrent layers for LSTMs); and attention heads \textbf{H}$\in\{2,4,8,16\}$ with Dim divisible by H.

Model-specific sliders included ARIMA’s $(p,d,q)$ orders, a moving-average window $m_a\in\{5,\dots,50\}$, PatchTST fully connected and head dropouts $fc_{dr},h_{dr}\in\{0.1,0.2,0.3,0.4,0.7\}$,  patch length and stride $p_t,str\in\{8,16,24\}$, FEDformer variant $v\in\{\text{Fourier},\text{Wavelets}\}$ with modes $m\in\{16,\dots,64\}$, and the DLinear \texttt{individual} flag $\in\{0,1\}$. The configuration yielding the lowest mean validation MSE for each model was retained for final BTCUSDT and ETHUSDT evaluation and is reported in Tables~\ref{tab:hyperparameters_btc} and~\ref{tab:hyperparameters_eth}.

\subsection{Evaluation Metrics}
We evaluated model performance on two fronts: forecasting accuracy and simulated trading utility. The trading strategy is a simple sign-based rule: go \textit{long} if predicted return $\hat{r}_{t+1} > 0$ and \textit{short} otherwise. The specific metrics are:
\begin{itemize}
    \item \textbf{Forecasting Accuracy:} Mean Squared Error (MSE, used as the training loss), RMSE, and Mean Absolute Error (MAE).
    \item \textbf{Trading Utility:} Total Return on Investment (ROI), Sharpe Ratio (risk-adjusted return), Maximum Drawdown, and Directional Accuracy (DA), defined as the percentage of correct sign predictions: $DA = \frac{1}{N} \sum_{t=1}^{N} \mathbf{1}_{\operatorname{sgn}(r_t) = \operatorname{sgn}(\hat{r}_t)}$.
\end{itemize}

\section{Results}

In this section, we present the empirical results of the Partial Multivariate Transformer (PMformer) against baseline models on Bitcoin (BTCUSDT) and Ethereum (ETHUSDT) datasets. Detailed metrics are provided in Table~\ref{tab:btc_eth_results}.

\begin{table*}[!ht]
    \centering
    \caption{Detailed comparison of 12 models for BTCUSDT and ETHUSDT, with full breakdown by group and metric type. Best results in each row are bolded.}
    \label{tab:btc_eth_results}
   
    {\small
    \setlength{\tabcolsep}{3.5pt}
   
    \begin{tabularx}{\textwidth}{l *{12}{>{\centering\arraybackslash}X}}
    \toprule
   
    & \multicolumn{1}{c}{\textbf{Partial-M.}} & \multicolumn{3}{c}{\textbf{Univariate}} & \multicolumn{8}{c}{\textbf{Multivariate}} \\
    \cmidrule(lr){2-2} \cmidrule(lr){3-5} \cmidrule(lr){6-13}
   
    & \rot{PMformer} & \rot{Naive Repeat} & \rot{ARIMA} & \rot{LSTM} & \rot{Autoformer} & \rot{FEDformer} & \rot{DLinear} & \rot{Informer} & \rot{iTransformer} & \rot{LSTM} & \rot{PatchTST} & \rot{Transformer} \\
    \midrule \midrule
    \textbf{BTCUSDT} \\
    \midrule
    \multicolumn{13}{l}{\textit{Statistical Metrics}} \\
    \midrule
    MSE ($\times10^{-4}$) & \textbf{6.5798} & 14.308 & 7.2527 & 6.5832 & 6.6929 & 6.6658 & 6.8792 & 6.6062
    & 6.7169 & 6.7070 & 7.0079 & 6.7367 \\
    RMSE ($\times10^{-2}$) & \textbf{2.5651} & 3.7826 & 2.6930 & 2.5658 & 2.5871 & 2.5818 & 2.6228 & 2.5702 & 2.5917 & 2.5898 & 2.6472 & 2.5955 \\
    MAE ($\times10^{-2}$) & 1.8919 & 2.8814 & 1.9409 & \textbf{1.825} & 1.9377 & 1.9109 & 1.9189 & 1.8514 & 1.9271 & 1.8333 & 1.936 & 1.9116 \\
    \midrule
    \multicolumn{13}{l}{\textit{Trading Metrics}} \\
    \midrule
    Total ROI (\%) & 20.62 & -0.34 & -0.36 & 2.52 & 24.5 & \textbf{38.08} & 1.6 & 2.52 & 12.96 & -0.80
    & 7.61 & 2.69 \\
    Sharpe Ratio& 3.83 & -0.28 & -0.26 & 1.63 & 4.00 & \textbf{4.54} & 1.56 & 1.63 & 3.29 & -1.63
    & 2.71 & 1.75 \\
    Max Drawdown (\%) & \textbf{-13.8} & -50.0 & -57.6 & -26.2 & -17.8 & -15.1 & -30.0 & -26.2 & -21.5 & -83.3 & -24.8 & -21.1 \\
    Direction Accuracy (\%) & \textbf{59.8} & 50.8 & 44.5 & 52.4 & 59 & 59.4 & 55.1 & 52.4 & 58.6 & 47.6 & 55.3 & 53.9 \\
    \midrule \midrule
    \textbf{ETHUSDT} \\
    \midrule
    \multicolumn{13}{l}{\textit{Statistical Metrics}} \\
    \midrule
    MSE ($\times10^{-4}$) & \textbf{9.6063} & 21.651 & 13.022 & 10.361 & 9.8831 & 11.087 & 10.573 & 9.7276
    & 10.559 & 9.7446 & 9.6872 & 10.172 \\
    RMSE ($\times10^{-2}$) & \textbf{3.0994} & 4.6530 & 3.6086 & 3.2189 & 3.1437 & 3.3297 & 3.2516 & 3.1189 & 3.2494 & 3.1216 & 3.1124 & 3.1893 \\
    MAE ($\times10^{-2}$) & \textbf{2.1711} & 3.4725 & 2.5199 & 2.3159 & 2.2702 & 2.4443 & 2.3354 & 2.191 & 2.3052 & 2.1973 & 2.2008 & 2.3127 \\
    \midrule
    \multicolumn{13}{l}{\textit{Trading Metrics}} \\
    \midrule
    Total ROI (\%) & -0.67 & 0.39 & -0.32 & -0.65 & \textbf{0.59} & -0.36 & 0.30 & -0.75 & -0.84 & -0.68
    & -0.43 & -0.64 \\
    Sharpe Ratio & -0.84 & 0.57 & -0.01 & -0.74 & \textbf{0.68} & -0.16 & 0.51 & -1.14 & -1.55 & -0.85
    & -0.28 & -0.76 \\
    Max Drawdown (\%) & -75.7 & -51.8 & -57.5 & -75.8 & -58.1 & -49.3 & \textbf{-44.0} & -78.6 & -86.7 & -75.8 & -62.6 & -79.9 \\
    Direction Accuracy (\%) & 47.9 & 51.8 & 48.3 & 47.1 & 49.3 & 49.6 & \textbf{53.1} & 48.0 & 47.5 & 47.8 & 48.3 & 45.1 \\
    \bottomrule
    \end{tabularx}}
\end{table*}

\subsection{Performance on BTCUSDT}

For BTCUSDT, PMformer achieved the highest statistical accuracy, posting the lowest MSE ($6.5798\times10^{-4}$) among all tested models. This result confirms the effectiveness of the partial-multivariate approach, which narrowly outperformed strong baselines including univariate LSTM (MSE $\approx6.583\times10^{-4}$) and Informer (MSE $\approx6.6\times10^{-4}$).

In simulated trading, this predictive accuracy translated into strong risk-adjusted performance. While FEDformer yielded a higher Total ROI (+38.1\%) and Sharpe Ratio (4.54), PMformer provided superior risk control, securing the lowest maximum drawdown (-13.8\%) and the highest directional accuracy (59.8\%). These results frame a clear trade-off: PMformer's predictions enable a stable, risk-averse strategy, whereas only one model generated higher but more volatile returns.

\subsection{Performance on ETHUSDT}
The ETHUSDT results mirror some of the trends seen with Bitcoin, but they also reveal asset-specific differences. PMformer again achieved the lowest prediction error (MSE $9.61\times10^{-4}$), narrowly outperforming PatchTST (MSE $9.69\times10^{-4}$). However, this statistical edge failed to generate a profit. The trading simulation based on PMformer's predictions resulted in a negative Total ROI (-0.67\%) and Sharpe Ratio (-0.84). This failure was not unique to PMformer; nearly all models incurred losses, with the best-performing model (Autoformer) achieving only a marginal +0.59\% ROI. For ETHUSDT, these findings powerfully demonstrate that superior forecasting accuracy, as measured by standard error metrics, was an insufficient condition for achieving financial utility in our simulation.

The stark contrast between BTC and ETH results underscores an important point: a model that is statistically accurate is not guaranteed to be financially effective. PMformer’s superior error metrics on ETH did not translate to gains - this suggests that Ethereum’s market had characteristics (perhaps higher noise or different key drivers) that our feature set and modeling approach did not fully capture for trading purposes. Thus, while PMformer confirmed expectations on the forecasting front (yielding the lowest errors for ETHUSDT as expected), it also provided a cautionary confirmation that prediction accuracy alone is not enough for successful trading, especially in a market where even accurate forecasts might be too small relative to transaction costs or volatility to exploit.

\section{Discussion}

This study investigated the effectiveness of the PMformer for cryptocurrency price forecasting, confirming our hypothesis that a model using a selected subset of features offers a compelling balance between information-scarce univariate and noise-prone full multivariate models.

A key finding is the asset-dependent nature of performance. For BTCUSDT, PMformer's statistical accuracy was highest, suggesting that a limited set of technical indicators provides a sufficiently strong signal. For ETHUSDT, the accuracy margin was smaller and, crucially, it did not translate to profitable trades. This divergence underscores that the optimal feature set is not universal and depends on the specific market dynamics of the asset.

Our analysis reveals a pronounced disconnect between statistical accuracy and practical trading utility. For both assets, the models with the lowest prediction errors were frequently not the most profitable. For ETHUSDT, nearly all models led to financial losses. This highlights a fundamental limitation of standard error metrics like MSE, which are indifferent to the direction of an error. In trading, correctly predicting the direction of a price move is often more critical than accurately predicting its magnitude.

\section{Conclusion}

In this paper, we present and evaluate a partial multivariate approach for cryptocurrency forecasting. Our findings demonstrate that using a targeted subset of features can effectively balance the trade-off between insufficient information and noise, leading to improved statistical forecasting accuracy. The primary contribution of this work is two-fold: it validates the potential of selective feature utilization in financial time series models and, more importantly, it highlights the critical divergence between standard statistical metrics and practical trading performance. These results underscore the complexity of financial markets and point toward necessary shifts in how forecasting models are evaluated and optimized.

\bibliographystyle{IEEEtran}
\bibliography{bibliography}

\end{document}